\documentclass[letter]{aa} 
\usepackage{graphicx}
\usepackage{natbib}
\bibpunct{(}{)}{;}{a}{}{,}
\newcommand{\COa}{$^{12}$CO}
\newcommand{\COb}{$^{13}$CO}
\newcommand{\CC}{$^{12}$C/$^{13}$C}
\newcommand{\Tef}{\mbox{$T_{\rm eff}$}}
\newcommand{\Vt}{\mbox{$V_{\rm t}$}}
\newcommand{\feh}{\mbox{[Fe/H]}}
\newcommand{\cn}{\mbox{[C/H]/[N/H]}}

\newcommand{\vunit}{\mbox{\,km\,s$^{-1}$}}
\newcommand{\mic}{\mbox{$\,\mu$m}}

\newcommand{\rs}{RS~Oph}

\begin{document}
   \title{Metallicity and effective temperature of the secondary of RS Oph}


   \author{Ya. V. Pavlenko
          \inst{1,2}
          \and
          A. Evans\inst{1}
          \and
          T. Kerr\inst{3}
          \and
      L. Yakovina\inst{2}
          \and
          C. E. Woodward\inst{4}
          \and
          D. Lynch\inst{5}
          \and
          R. Rudy\inst{5}
      \and
      R. L. Pearson\inst{5,6}
      and
      R. W. Russell\inst{5}}

   \offprints{Ya. V. Pavlenko}

   \institute{  Astrophysics Group, Keele University, Keele, Staffordshire, ST5 5BG, UK
             (\email{ae@astro.keele.ac.uk})
       \and
   Main Astronomical Observatory of the National Academy of Sciences
              of Ukraine,  27 Zabolonnoho, Kyiv-127, 03680 Ukraine
              (\email{yp@mao.kiev.ua, yakovina@mao.kiev.ua})
         \and
             Joint Astronomy Centre, 660 N. A'ohoku Place, University Park,
             Hilo, Hawaii 96720, USA \\ (\email{tkerr@jach.hawaii.edu})
         \and
             Department of Astronomy, 116 Church St. SE, University of Minnesota,
             Minneapolis, MN 55455, USA (\email{chelsea@astro.umn.edu})
         \and
             The Aerospace Corporation, Mail Stop M2-266, PO Box 92957, Los Angeles,
             CA 90009-29957, USA (\email{David.K.Lynch@aero.org,
         Richard.J.Rudy@aero.org, Richard.L.Pearson@aero.org, \\ 
Ray.W.Russell@aero.org)}
         \and Department of Astronomy, Brigham Young University,  N283 ESC
             Provo, UT  84602, USA}

   \date{Received\ldots\ldots\ldots; accepted\ldots\ldots\ldots}

  \abstract
   {The recurrent nova {\rs}iuchi undergoes nova eruptions every $\sim10-20$~years 
    as a result of thermonuclear runaway on the surface of a white dwarf close
    to the Chandrasekhar limit. Both the progress of the eruption, and its aftermath,
    depend on the (poorly known) composition of the red giant in the \rs\ system.}
   {Our aim is to understand better the effect of the giant secondary on the
    recurrent nova eruption.}
   {Synthetic spectra were computed for a grid of M-giant model atmospheres
    having a range of effective temperatures
    $3200 < \Tef < 4400$~K, gravities 0 $<$ log g $<$ 1 and
    abundances -4 $<$[Fe/H] $<$ 0.5, and fit to
    infrared spectra of RS Oph as it returned to quiescence after its 2006 eruption.
    We have modelled the infrared spectrum in the range $1.4-2.5$\mic\
    to determine metallicity and effective temperature of the red giant.}
    {We find $\mbox{\Tef} = 4100 \pm 100$~K, $\log g = 0.0 \pm 0.5$, 
     $\mbox{[Fe/H]} = 0.0 \pm 0.5$,
    $\mbox{[C/H]} = -0.8 \pm 0.2$, $\mbox{[N/H]} = +0.6 \pm 0.3$ in the 
    atmosphere of the secondary, and demonstrate
    that that inclusion of some dust `veiling' in the spectra
    cannot improve our fits. }
   {}


   \maketitle
%

\section{Introduction}

The recurrent nova (RN) {\rs}iuchi is a binary system consisting of a red 
giant (RG) star and a white dwarf (WD) with mass near the Chandrasekhar 
limit \citep[see][~and references therein]{shore,fekel}. \rs\ is the best 
studied of RNe, and is known to have undergone at least five eruptions, in 
1898, 1933, 1958, 1967 and 1985. Each eruption displays very similar 
visual light curves \citep[e.g.,][]{rosino}. The most recent outburst of 
\rs\ was discovered in 2006 February 12.829 UT \citep[][~this defines our 
time origin]{hirosawa}, and was the subject of an intensive, 
multi-wavelength observational campaign, from the radio to x-rays 
\citep{e2006}. 

It is likely that, unlike {\em classical} novae, in which some of the WD is 
stripped away in the nova eruption, resulting in a secular decline in the 
WD mass, the mass of the WD in \rs\ may be {\em increasing}. If this is 
the case, the mass of the WD in \rs\ may eventually reach the 
Chandrasekhar limit and explode as a Type Ia supernova 
\citep{starrfield,wood}. However this remains a matter of considerable 
debate \citep[e.g.,][]{ness08}.

Despite a long history of observations of \rs, our knowledge of the 
secondary star is surprisingly sparse. The secondary known to be a RG, 
({\it SIMBAD}\footnote{http://simbad.u-strasbg.fr/Simbad} spectral type 
M2IIIpe+) yet little is known about photospheric abundances. 
\cite{scott} claimed some evidence of a deficit of carbon in the 
secondary, while \cite{wallerstein} found a ``rather small excess'' of 
metals. 

A knowledge of abundances in the atmosphere of the RG in the \rs\ system 
is of crucial importance for two reasons. First, in contrast to classical 
novae, it seems that only material accreted by the WD from the RG takes 
part in the thermonuclear runaway (TNR) that leads to the RN eruption. A 
complete understanding of the TNR requires knowledge of the composition of 
material deposited on the WD; moreover, knowing the composition of the 
material accreted by the WD will make it possible to predict, with some 
confidence, the composition of the ejected material, without the 
complication of knowing (or guessing) the amount of material dredged up 
from the WD. Second, the material ejected in the RN eruption runs into the 
RG wind \citep{BK}, which is shocked, causing the gas to emit strongly at 
x-ray \citep{bode06,ness07}, infrared 
\citep[IR;][]{das06,evans07a,evans07b} and radio wavelengths 
\citep{obrien}. Indeed x-ray observations of the 2006 eruption 
\citep{ness08} show complex, evolving line emission, the interpretation of 
which requires knowledge of the composition of the RG wind, which is 
poorly characterised. 

Here we report the results of a preliminary attempt to rigorously determine 
some of the parameters of the RG in the \rs\ system by modelling its IR 
spectrum after the 2006 eruption had subsided. Special attention is paid 
to the abundances of carbon and nitrogen, and to the isotopic ratio \CC, 
which are primary indicators of the state of stellar evolution. 

From the known orbit \citep{fekel} we find that the \rs\ system was close 
to quadrature during our observation, so that the WD was illuminating half of 
the observed photospheric hemisphere of the RG. Further papers will 
augment the results presented here, and will address the issue of the 
irradiation of the RG by the WD. 

\section{Observational data}
\label{ukirt}

\rs\ was observed in the $I\!J\!H\!K$ bands with the UIST instrument 
($\lambda/\Delta\lambda \simeq 1500$) on the United Kingdom Infrared 
Telescope (UKIRT) on 2006 August 25 (day 193.71) and on September 18 (day 
212.9). While the data cover the wavelength range 0.87--2.51\mic\ only 
data in the 1.4--2.5\mic\ range are considered here. First order sky 
subtraction was achieved by nodding along the slit; HR6493 was used to 
remove telluric features and for flux calibration. Wavelength calibration 
used an argon arc, and is accurate to $\pm0.0003$\mic\ in the $H\!K$ bands; 
further observational details may be found in \cite{evans07a}. The 
data were dereddened using $E(B-V)=0.73$ \citep{snijders}.

We note that data at the edge of the atmospheric window may be less reliable
than elsewhere due to inadequate atmospheric cancellation. While this has
minimal effect on the most pertinent diagnostic (molecular) features this
may impact on our fit at the window edges.

\section{Procedure}
\label{procedure}
\subsection{Origin of CO bands}

We are confident that the CO and CN bands originate in the atmosphere of 
the RG and may be used as a diagnostic of photospheric conditions. Some 
classical novae are known to display CO emission \citep[e.g.,][]{evans05} 
which may cause `veiling' of the photospheric CO but the environment of 
\rs\ during its eruption would never have been conducive to molecule 
formation. Furthermore there is little or no change in the molecular bands 
between the August and September observations, so we are proceed on the 
assumption that these features arise in the RG atmosphere. 

\subsection{Model atmospheres and synthetic spectra}
\label{MA}

We compute plane-parallel model atmospheres of evolved stars in LTE, with 
no energy divergence, using the SAM12 program \citep{P93}, which is a 
modification of ATLAS12 \citep{kurucz99}. Chemical equilibrium is computed 
for molecular species assuming LTE. The opacity sampling approach 
\citep{sneden} is used to account for absorption of atoms, ions and 
molecules \citep[for more details see][]{P93}. The 1-D convection mixing 
length theory modified by \cite{kurucz99} in ATLAS12 was used to account 
for convection. The computed model atmospheres are available on the 
web\footnote{ ftp://ftp.mao.kiev.ua/pub/users/yp/RS.Oph}. 

Synthetic spectra are calculated with the WITA6 program 
\citep{pavlenko00}, using the same approximations and opacities as SAM12. 
To compute synthetic spectra we use line lists from 
\citet[][~TiO]{plez89}, \citet[][CO]{goorvitch}, \citet[][VALD]{kupka}, 
\citet[][CN]{kurucz93}, and \citet[][H$_2$O]{BT2}. The shape of each 
molecular or atomic line is determined using the Voigt function. Damping 
constants are taken from line databases, or computed using Unsold's 
(\citeyear{unsold}) approach. A wavelength step $\Delta\lambda$ = 0.5 \AA\ 
is employed in the synthetic spectra computations. 

In our computations we adopt a microturbulent velocity \Vt = 3\vunit. This 
value is somewhat higher than the $\Vt \sim 2$\vunit\ found in the 
atmospheres of RGs \citep[see][]{foy}. However a higher value may be more 
realistic for the secondary of a RN. Unfortunately our spectra are not 
suitable for a more accurate determination of \Vt; we will further 
investigate this in  forthcoming papers. 

\subsection{\label{veil}Fits to observed spectra}

To determine the best fit parameters, we compare the observed fluxes
$F_{\nu}$ with the computed fluxes $F_{\nu}^x$ following the scheme of
\cite{pav-jones}. We adopt a Gaussian profile (FWHM = 0.002 \mic) to model
line broadening. We then find the minima of the 3D function
$$S(f_{\rm s}, f_{\rm h}, f_{\rm g}) =
   \sum_\nu \left ( F_{\nu} - F_{\nu}^x \right )^2  , $$
where $F_{\nu}$ and $F_{\nu}^x$ are the observed and computed spectra
respectively, and $f_{\rm s}$, $f_{\rm h}$, $f_{\rm g}$ are the
wavelength shift, the normalisation factor, and the profile
broadening parameter, respectively. The model parameters
are determined by minimising $S$ for every computed
spectrum, and from the grid of the better solutions for a given set
of abundances and/or other parameters (microturbulent velocity,
effective temperature, isotopic ratios, etc.), we choose the
best-fitting solution.

For many evolved stars, some of the flux, particularly at longer 
wavelengths, arises from dust in the circumstellar environment. On the 
other hand any inadequate atmospheric cancellation (see 
Section~\ref{ukirt}) can mimic the presence of circumstellar dust. Either 
of these effects provides a source of `veiling' in the observed spectra, 
such that $F_{\rm total} = F_{\rm atmos}+a_0$, where $a_0$ is the flux 
formed outside the stellar photosphere. In this case we should minimise 
$F_{\rm total}-F_{\rm obs}$, whilst considering $F_{\rm envelope}$ as an 
additional parameter (see \citealt{pav-trg} and \citealt{pav04} for more 
details). In the present paper we refer to both of these effects as 
`veiling'. 

\subsection{Algorithm to determine best fit}

We carry out the process in several steps, as follows:

\begin{enumerate}
\item we compute a small grid of model atmospheres in the range of 
metallicities $\mbox{[Fe/H]} = 0.5, 0, -1, -2, -3, -4$, effective 
temperatures $\mbox{\Tef} = 3400, 3600, 3800, 4000$~K and $\log g = 0, 
0.5, 1$; 

\item the main absorption features in the observed spectrum were identified
(excluding the emission lines arising in the shocked wind and ejecta);

\item for these models we compute synthetic spectra;

\item after determining the basic parameters of \rs\ we compute further
grids of synthetic spectra, but incorporate at this stage a range of possible 
non-solar abundances for carbon and nitrogen;

\item the best fits were found following the schemes described in
Section~\ref{veil}, i.e. we obtained solutions with and without veiling;

\item the best fit was determined from the both non-veiled and veiled solutions.

\end{enumerate}

\section{Results}
\label{results}

  \begin{figure}
   \centering
   \includegraphics[width=70mm]{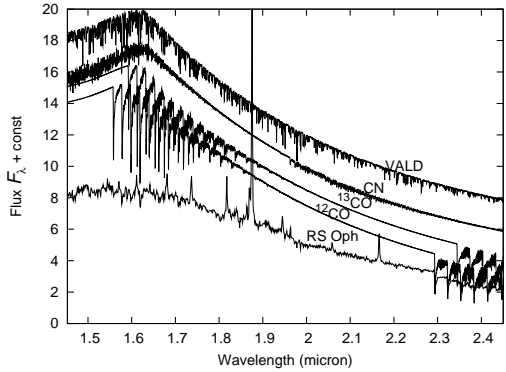}
      \caption{Contribution of different species to the formation of the
      observed spectrum of \rs\ (bottom plot), which still displays emission
      lines following the eruption.
      Computed spectra, showing contributions of $^{12}$CO, $^{13}$CO and CN
      are shifted vertically for better presentation}
         \label{_1i}
   \end{figure}

   \begin{figure}
   \centering
   \includegraphics[width=70mm,height=60mm]{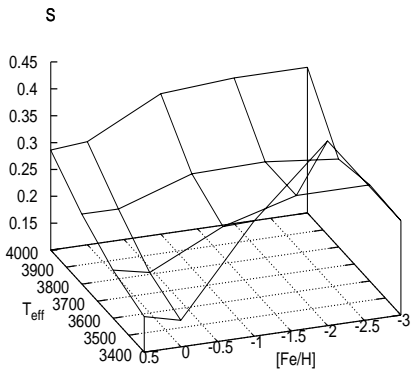}
      \caption{Determination of the best fit from the family of
      solutions obtained for model atmospheres of different \Tef\ and [Fe/H].}
         \label{_ddd}
   \end{figure}
   
      \begin{figure*}
   \centering
   \includegraphics[width=140mm]{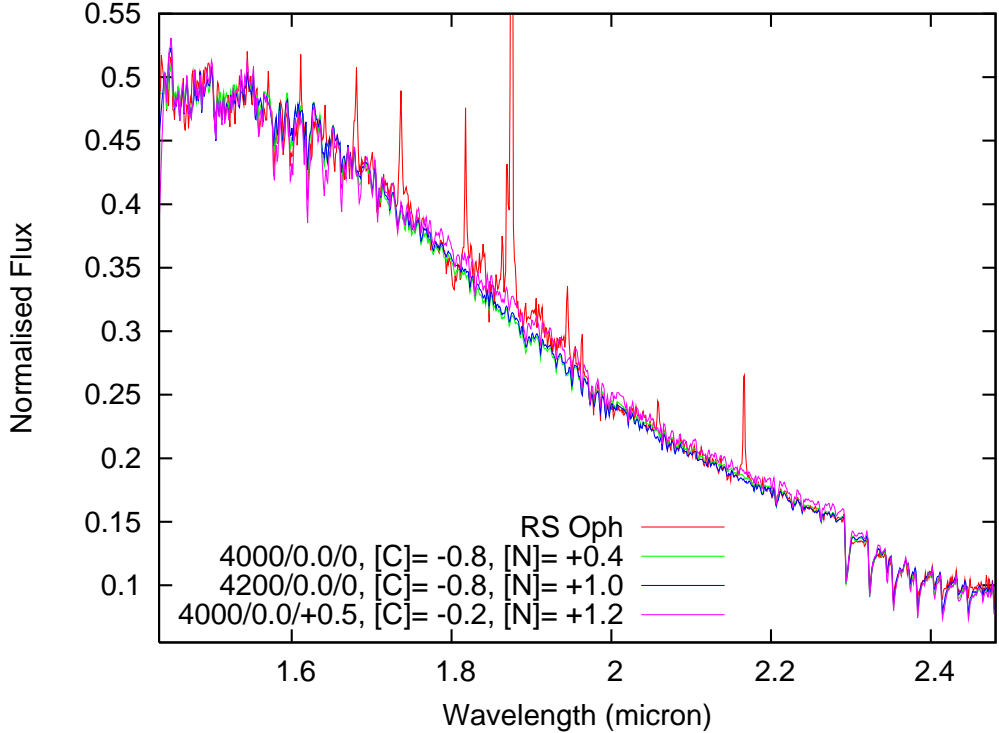}
      \caption{Three best fits to the observed spectrum of RS Oph
      for the ``veiling free'' models shown in the Table \ref{_t1}.}
       \label{_seds}
   \end{figure*}

\subsection{Fits to the SED of \rs: models with different metallicities}

The main features in the photospheric spectrum of \rs\ are absorption by 
\COa\ and \COb\ around 1.6 and 2.3\mic\ respectively, where the first and 
second overtone ($\Delta \upsilon = 2$ and 3 respectively) bands of CO are 
located (see Fig.~\ref{_1i}). Some features at $\lambda <$ 1.6\mic\ are 
due to absorption by CN bands.

The slopes of the spectral energy distribution (SED) and the intensity of
molecular bands in the modelled spectra depend on both \Tef\ and [Fe/H].
This allows us to refine both parameters from analysis of the best fits
of the synthetic spectra to the observed spectrum of \rs. In Fig.~\ref{_ddd}
we show the dependence of the best fit parameter $S$ on \Tef\ and on [Fe/H].
Here the range of metallicities is restricted to the range [--2\ldots +0.5];
for lower metallicities the minimisation does not produce any meaningful
fits because the features in the computed spectra become too shallow.
The computations in Fig.~\ref{_ddd} were carried out for the veiling-free case,
and the abundances of carbon and nitrogen were scaled by the factor [Fe/H].

The best fit over the wavelength range 1.506--2.490\mic\ has $S = 0.1565 
\pm 0.0005$ for $\mbox{\Tef} =3400 \pm 100$~K, $\log g = 0.0 \pm 0.5$ and 
$\mbox{[Fe/H]} = 0.0 \pm 0.5$ We note that our best solution is consistent 
with normal (solar) metallicities (see Anders \& Grevesse 1989). 
However over the wavelength range 
1.491--2.490\mic\ (which includes the CN bands at the blue end of the 
spectral range), we obtain a better solution, with $S$ = 0.2066 $\pm$ 
0.0006 for a model atmosphere with the same \Tef = 3400 $\pm$ 100~K, and 
[Fe/H] = 0.0 $\pm$ 0.5, but with $\log g = 1 \pm$ 0.5 (note that the value 
of $S$ depends on the wavelength range over which the comparison between 
observed and computed spectra is made). However, further analysis 
(Section 4.2) shows that the CO and CN bands in our theoretical spectra do 
not have the proper intensities for these parameters if simultaneous fits 
are attempted over a broad wavelength range.

\subsection{Fits to the SED of \rs: models with different abundances of carbon
and nitrogen}

Numerical experiments showed that, in order to get acceptable fits
of both CO and CN bands, the carbon abundance must be reduced,
while the nitrogen abundance must be increased. To determine the
appropriate abundances of carbon and nitrogen we compute the
second set of synthetic spectra varying 
abundances of carbon log N(C) and nitrogen log N(N)
over a wide range.

Computations were carried out for the case of (i)~``normal metallicities'', 
i.e. all abundances other than N and O were taken to be solar, [Fe/H] = 
0.0, and (ii)~slightly enhanced metallicities, [Fe/H] = +0.5. For model 
atmospheres having solar abundances we performed computations for  
$-1.2 < \mbox{[C/H]} < -0.2$ and $0 <  \mbox{[N/H]} < +1.2$. 
Sets of variable carbon and nitrogen 
abundances were adopted with a step of 0.2~dex. We note that the adopted 
variations of the C and N abundances have minimal effect on the 
temperature structure of the RG model atmosphere for $\mbox{\Tef}$ in the 
range 3600--4400~K. With these refined abundances of carbon and nitrogen 
we carried out the entire set of computations and comparison. 

\begin{figure}
   \centering
   \includegraphics[width=70mm]{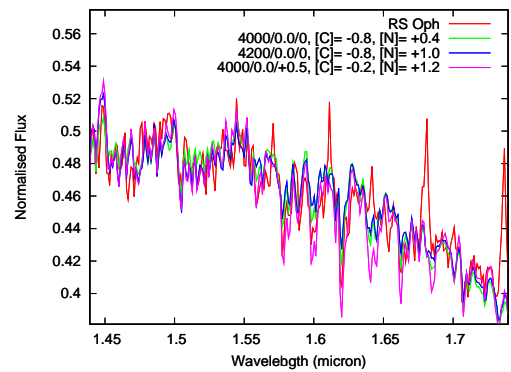}
      \caption{Detail from Fig.~\ref{_seds} around 1.6\mic;
      main absorption features are due to CN and CO ($\Delta v$ = 3) bands. }
         \label{_sedsl}
   \end{figure}

\begin{figure}
    \centering
   \includegraphics[width=70mm]{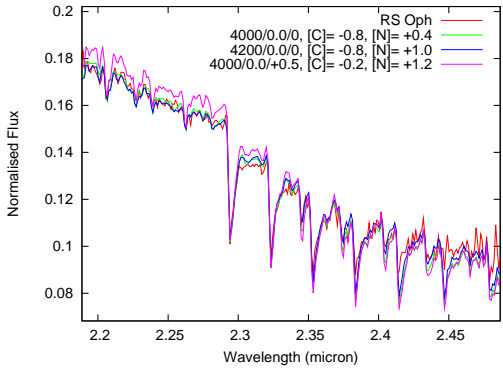}
      \caption{Detail from Fig.~\ref{_seds} around 2.3\mic; main absorption
features here are due to CO ($\Delta v$ = 2) bands.}
         \label{_sedsr}
   \end{figure}

Better fits to the observed SEDs were obtained for model atmospheres with 
\Tef/$\log g$/[Fe/H] = 4000/0.0/0.0, 4000/1.0/+0.5, 4200/0.0/0.0 with $S$ 
= 5.176 $\pm$ 0.031, 5.110 $\pm$ 0.031, 5.130 $\pm$ 0.031, 
Figs.~\ref{_seds}, \ref{_sedsl}, \ref{_sedsr}, respectively. Formally the 
best solution we obtain is for the 4000/1.0/+0.5 set of parameters. However, as 
illustrated in Fig.~\ref{_sedsr}: 

\begin{itemize}
\item the fit to the observed slope at $\lambda > 2$\mic\ is better
with model spectra computed with the 4000/0.0/0.0 and 4200/0.0/0.0 model
atmospheres;

\item a few of the sufficiently strong atomic lines computed for the metal-rich
model atmosphere are too strong by comparison with observations.
\end{itemize}

\noindent Both these factors provide good constrains for choosing the best
solution, which is provided by two model atmospheres 4000/0.0/0.0 or
4200/0.0/0.0, and with $\mbox{[C/H]}=-0.8 \pm 0.2$, $\mbox{[N/H]}=+0.6 \pm 0.3$.

Emission measure distribution analysis and APEC modelling of line 
fluxes derived {\em Chandra} and {\em XMM-Newton} spectra obtained over a 
period 240 days post-outburst \citep{ness08} indicates that the 
metallicity of the ejecta is $\approx 0.5$ solar, with N to be 
overabundant and Fe to be underabundant relative to O. The N 
overabundance derived by \cite{ness08} suggests that the material accreted 
by the WD from the RG secondary is N-enhanced, in qualitative 
agreement with our models.

\begin{table*}
 \centering
 \caption{\label{_t1}  Log abundances of carbon and nitrogen obtained from the best fit
 of the infrared spectra of RS Oph}
 \begin{tabular}{ccccccccc}
 \hline
 \noalign{\smallskip}

\multicolumn{9}{c}{Dusty-free case} \\

\hline
\noalign{\smallskip}

\Tef (K) & \multicolumn{2}{c}{3800}    & \multicolumn{2}{c}{4000} & \multicolumn{2}{c}{4200} & \multicolumn{2}{c}{4400} \\
\hline
\noalign{\smallskip}

\feh &  0.0 &   0.5&      0.0 & 0.5   &       0.0 & 0.5 &       0.0 &  0.5 \\
$S$*100 & 6.403& 5.667&    5.176 &5.110  &    5.130 &6.019 &    5.393  &9.355 \\
log g &   0.0&   0.5&      0.0 &   1.0 &       0.0 & 0.5 &      1.0    &  1.0  \\
\cn&-1./0& -0.5/+0.4& -0.8/+0.4& -0.2/+1.2&-0.8/+0.9&-0.2/+0.2 &-0.2/+0.8 &-0.2/+1.2 \\
\\

\hline
\hline
\noalign{\smallskip}
\multicolumn{9}{c}{ ``Veiled'' model} \\

\hline
\noalign{\smallskip}

\Tef (K) & \multicolumn{2}{c}{3800} & \multicolumn{2}{c}{4000} & \multicolumn{2}{c}{4200} & \multicolumn{2}{c}{4400} \\
\hline
\noalign{\smallskip}
\feh &  0.0  & 0.5   &   0.0  &  0.5     &      0.0  &  0.5   &      0.0   &0.5      \\
$S$*100&  9.176 &6.929&   5.738&   6.149  &      5.374&   5.900&       5.574&  7.653  \\
log g  & 0.0  & 0.0   &   0.0  &   1.0      &      0.5  &   1.0    &        1.0   &  0.5    \\
\cn& -0.8/0&-0.8/+0.2& -0.6/+0.8& -0.4/+1.0  &    -0.4/+0.4&-0.3/+1.2&   -0.2/+0.6 &-0.2/+1.2   \\
a$_0$  &   0.02 & 0.02&    0.12&   0.02   &       0.02&   0.02 &     0.02  &    0.02   \\

\hline
\hline
\noalign{\smallskip}
\end{tabular}
\centering
\end{table*}

The deduced effective temperature \Tef\ (4000-4200~K) of RS Oph corresponds 
well with the upper limit of the conventional mean values for spectral 
classes M0--M2 (see http://www.astro.umd.edu/\~{ }roger/pubs/mtemps.ps). 
Perhaps, this \Tef\ is a consequence of irradiating of the 
photosphere of RS Oph by WD. In general this range of \Tef\ 
corroborates our expectation that the effect of irradiation of atmosphere 
of RS Oph should not significantly affect the IR spectra. The effect of 
irradiation should be more critical for the optical part of the spectrum, 
especially its blue and UV regions. 

The inclusion of even a small amount ``veiling'' (see 
Section~\ref{veil}) reduces the quality of the fits (see Table \ref{_t1}). 
This result implies that at the time of our IR observations the 
contribution of any dust envelope \citep{evans07b} around \rs\ 
was negligible.

\subsection{Fits to the CN and CO bands}

We obtain good fits to the CN bands around 1.6\mic, and to the CO bands 
around 1.6 (Fig.~\ref{_sedsl})  and 2.3\mic\ (Fig.~\ref{_sedsr}). In the 
2.3\mic\ range both \COa\ and \COb\ bands are observed and in principle, 
we can use these to determine the isotopic ratio \CC. However, due to the 
low spectral resolution of our data, we can only infer a lower limit 
on the \CC\ ratio in the RG photosphere, $\mbox{\CC} > 10$. The model 
spectra shown in Figs.~\ref{_sedsl}, \ref{_sedsr} were computed for 
\CC $= 14$, and the general picture is not changed for a somewhat lower 
value, $\mbox{\CC} = 9$. We will report better estimates of \CC\ in 
forthcoming papers. 

\section{\label{_D} Discussion and conclusion}

Our results are obtained in the `classical' approach, using plane-parallel 
model atmospheres with no sinks or sources of energy. However, at the time 
of our observation, the IR spectrum displayed strong H and He lines from 
the shocked wind and ejecta \citep{evans07b}. \rs\ is (following outburst) 
known to be a strong source of x-radiation and in principle, the flux of 
high energy photons could change the outermost layers of RG atmosphere. 
However, the IR flux originates deep in the atmosphere of the RG, so that 
a thick slab of cool material lies above the IR photosphere: high energy 
photons cannot penetrate these layers and are not likely to affect our 
general conclusions. We will be assessing the effects of irradiation 
as the RG is observed with the WD in different orbital configurations.

There is evidence for a modest deficit of carbon, and overabundance 
of nitrogen, in the atmosphere of the RG. It may be that carbon was 
converted to nitrogen in the CN cycle (see \citealt{sneden2000}) during 
the previous evolution of the secondary. Alternatively, there may have 
been some pollution of the RG surface by products of former   
eruptions. Nevertheless, the C and N abundance, together with the \CC\ 
ratio when it becomes available, will provide new constraints for 
theoretical models of the TNR and its aftermath. 

Furthermore, determination of the oxygen abundance in the atmosphere
of the RG will be of crucial importance for two reasons: first, oxygen
cannot be formed in the interiors of low and intermediate mass stars
and second, oxygen may have been produced in past RN eruptions.
In any case, the determination of the oxygen abundance
should clarify the origin of the carbon and nitrogen abundances.

Our UKIRT \rs\ programme is ongoing and in future papers we will address the
effects of irradiation, the \CC\ ratio.


\begin{acknowledgements}

The United Kingdom Infrared Telescope is operated by the Joint Astronomy
Centre on behalf of the U.K. Science and Technology Facilities Council
(STFC). This work used the research computing facilities at the                        
Centre for Astrophysics Research, University of Hertfordshire.
This work was supported by an International Joint Project Grant
from the UK Royal Society and the ``Microcosmophysics'' program of
the National Academy of Sciences and Space Agency of Ukraine.

\end{acknowledgements}

\end{document}